# Point and interval estimation of exposure effects and interaction between the exposures based on logistic model for observational studies

Xiaoqin Wang[1], Weimin Ye[2] and Li Yin[2,*]

[1]Department of Electronics, Mathematics and Natural Sciences, University of Gävle, SE-801 76, Gävle, Sweden

[2]Department of Medical Epidemiology and Biostatistics, Karolinska Institute, Box 281, SE-171 77, Stockholm, Sweden

[*]Corresponding author: Email: li.yin@ki.se

## Summary

In observational studies with dichotomous outcome of a population, researchers need to present the effects of exposures and interaction between the exposures jointly in order to learn the relationship between the exposure effects and the interaction. In this article we study point and interval estimation of exposure effects and the interaction based on logistic model, where the exposure effects are measured by risk differences while the interaction is measured by difference between risk differences. Using approximate normal distribution of the maximum-likelihood (ML) estimate of the model parameters, we obtain approximate non-normal distribution of the ML estimate of the exposure effects and the interaction. Using the obtained distribution, we obtain point estimate and confidence region of (exposure effect, interaction) as well as point estimate and confidence interval of the interaction when the ML estimate of an exposure effect falls into specified range. Our maximum-likelihood-based approach provides a simple but reliable method of interval estimation of exposure effects and the interaction.

## 1. Introduction

Suppose one conducts a randomized trial to investigate the effect of two exposures $z_1$ and $z_2$ on an outcome $y$ of certain population, where $z_1$, $z_2$ and $y$ are all dichotomous, namely, $z_1 = 0, 1$, $z_2 = 0, 1$ and $y = 0, 1$. In the randomized trial, covariates are essentially unassociated with the exposures $(z_1, z_2)$ and thus are not confounders. Oftentimes, one measures the effect of $z_1$ when $z_2 = 0$ by the risk difference

$$\text{TE1} = \text{pr}(y = 1 \mid z_1 = 1, z_2 = 0) - \text{pr}(y = 1 \mid z_1 = 0, z_2 = 0)$$

and the interaction between $z_2$ and $z_1$ by the difference between risk differences [1--5]

$$\text{INT} = \{\text{pr}(y = 1 \mid z_1 = 1, z_2 = 1) - \text{pr}(y = 1 \mid z_1 = 0, z_2 = 1)\} - \text{TE1}.$$

Likewise, one measures the effect of $z_2$ when $z_1 = 0$ by the risk difference

$$\text{TE2} = \text{pr}(y = 1 \mid z_1 = 0, z_2 = 1) - \text{pr}(y = 1 \mid z_1 = 0, z_2 = 0)$$

and obtains another expression of the above interaction as

$$\text{INT} = \{\text{pr}(y = 1 \mid z_1 = 1, z_2 = 1) - \text{pr}(y = 1 \mid z_1 = 1, z_2 = 0)\} - \text{TE2}.$$

INT is also called the biological interaction because its expression in terms of risk differences has a close relationship with some well-known classification of biological mechanisms [3, 4]. When presenting INT, one also presents TE1 and TE2 in order to learn the significance of INT relative to TE1 and TE2. For instance, INT = 0.05 has rather different significance relative to TE1 = 0.01 versus TE1 = 0.5.

Now suppose one conducts an observational study in which one also wishes to study TE1, TE2 and INT. In the observational study, there exist confounders, which are associated with the outcome $y$ and the exposures $(z_1, z_2)$. The most common model to adjust for confounding covariates in estimating the effect of $(z_1, z_2)$ is logistic model. Most often, logistic model is used to obtain point estimate and confidence interval of odds ratio as measure of the exposure effects as well as point estimate and confidence interval of ratio of odds ratios as measure of the interaction. In recent years, logistic model is also used to obtain point estimate and confidence interval of TE1 or TE2 as measures of exposure effects [6--9], and in rare cases, of INT as measures of interaction [10]. Noticeably, TE1, TE2 and INT are more interpretable than odds ratio and ratio of odds ratios.

In their studies, confidence interval of TE1, TE2 or INT is obtained by the normal approximation method and the parametric / non-parametric bootstrap method; the normal approximation method is based on approximate variance of the ML estimate of TE1, TE2 or INT; the bootstrap method generates bootstrap samples and then uses the samples to obtain the bootstrap distribution of TE1, TE2 or INT and then the bootstrap confidence interval. However, little is seen in the literature that reports TE1, TE2 and INT jointly, e.g. confidence region of (TE1 or TE2, INT), which takes into account the correlation between the estimates of (TE1 or TE2, INT).

In this article, we use logistic model to obtain point and interval estimate of (TE1 or TE2, INT). Instead of deriving the covariance matrix of the two-dimensional ML estimate of (TE1 or TE2, INT) or using the bootstrap method, we generate approximate distribution of the ML estimate of (TE1 or TE2, INT) and then use the distribution to obtain the interval estimate of (TE1 or TE2, INT). We present our method by studying the effect of (hospital type, cancer stage) on cancer survival.

## 2. Effect of (hospital type, cancer stage) on cancer survival in Sweden
### 2.1 Medical background and the data

In cancer treatment, an important question is which type of hospitals, small or large, is superior to treatment of cancers of early or advanced stage, where the size of a hospital is determined by the number of cancer patients treated there. In an observational study, researchers studied the effect of (hospital type, cancer stage) on one-year survival of cardia cancer patients [11]. Cardia cancer is highly malignant with bad prognosis and its one-year survival is a good measure of the performance of hospital type and the impact of cancer stage in treating such cancers. The data was collected between 1988 and 1995 on 150 cardia cancer patients treated in hospitals located in central and northern Sweden.

The hospitals were categorized into two types: large type ($z_1 = 1$) when treating more than 10 patients during the period 1988-1995, and small type ($z_1 = 0$) when treating less than or equal to 10 patients. Cancer stages were categorized into two categories: advanced stage ($z_2 = 1$) when the eight-level stage index took values larger than or equal to five, and early stage ($z_2 = 0$) when the index took values smaller than five. Then the exposures in this study were $\boldsymbol{z} = (z_1, z_2) =$ (hospital type, cancer stage). The outcome of a patient was successful ($y = 1$) versus unsuccessful ($y = 0$) survival for one year after diagnosis.

In addition to the exposure $\boldsymbol{z}$, possible confounders were documented: age ($x_1$), gender ($x_2$), and geographic area ($x_3$). Age was continuous, but gender and geographic area were categorical. Let $x_2 = 1$ indicate male and $x_2 = 0$ female of the gender. Geographic area was categorized into urban ($x_3 = 1$) *versus* rural ($x_3 = 0$). Let $\boldsymbol{x} = (x_1, x_2, x_3)$ be the set of the documented covariates. The descriptive statistics of these covariates and exposures are given in Table 1. The complete data of the study is given as supplementary material.

## 2.2  Exposure effects and interaction between the exposures

Let $y(z_1, z_2)$ be the potential outcome of each patient in the population under exposure $(z_1, z_2)$, where $z_1 = 1, 0$ indicates large versus small hospital types while $z_2 = 1, 0$ indicates cancers of advanced versus early stages, see [12--16] for the framework of causal inference. We use $\text{pr}\{y(z_1, z_2) = 1\}$ to denote the risk of $y(z_1, z_2) = 1$ of patient under exposure $(z_1, z_2)$. Then the effect of the hospital type $z_1$ at the early cancer stage $z_2 = 0$ is measured by the risk difference

$$\text{TE1} = \text{pr}\{y(1, 0) = 1\} - \text{pr}\{y(0, 0) = 1\}.$$

The effect of the cancer stage $z_2$ at the small hospital type $z_1 = 0$ is measured by the risk difference

$$\text{TE2} = \text{pr}\{y(0, 1) = 1\} - \text{pr}\{y(0, 0) = 1\}.$$

The interaction between $z_1$ and $z_2$ is measured by the difference between risk differences

$$\text{INT} = [\text{pr}\{y(1, 1) = 1\} - \text{pr}\{y(0, 1) = 1\}] - [\text{pr}\{y(1, 0) = 1\} - \text{pr}\{y(0, 0) = 1\}]$$
$$= [\text{pr}\{y(1, 1) = 1\} - \text{pr}\{y(0, 1) = 1\}] - \text{TE1},$$

or equivalently,

$$\text{INT} = [\text{pr}\{y(1, 1) = 1\} - \text{pr}\{y(1, 0)\} = 1] - [\text{pr}\{y(0, 1) = 1\} - \text{pr}\{y(0, 0) = 1\}]$$
$$= [\text{pr}\{y(1, 1) = 1\} - \text{pr}\{y(1, 0)\} = 1] - \text{TE2}.$$

In some medical applications, one aims at the effect of $z_1$ in stratum $z_2 = 1, 0$, namely,

$$\text{TE}(z_2) = \text{pr}\{y(z_1 = 1) = 1 | z_2\} - \text{pr}\{y(z_1 = 0) = 1 | z_2\},$$

where $y(z_1)$ is the potential outcome of each patient in stratum $z_2$ under exposure $z_1$, and then considers the modification of the effect of $z_1$ by $z_2$, namely,

$$\text{EM} = \text{TE}(z_2 = 1) - \text{TE}(z_2 = 0).$$

However, EM compares the effects of $z_1$ on different strata $z_2 = 1, 0$ and does not have causal interpretation. Thus one cannot use EM to address the question of the illustrative example described in the previous subsection. We have similar situation for the modification of the effect of $z_2$ by $z_1$.

Because we can only observe potential outcome of a patient under one of the four exposures $(z_1, z_2) = (0,0), (0,1), (1,0), (1,1)$, we need certain assumption to allow for estimation of TE1, TE2 and INT [12--16]. In the medical context of this study, it is reasonable to assume that there is no other confounder than the documented covariates $x = (x_1, x_2, x_3)$. We denote the risk of $y = 1$ in stratum $(z_1, z_2, x)$ by $\text{pr}(y = 1 \mid z_1, z_2, x)$. Then the assumption states

$$\text{pr}\{y(z_1, z_2) = 1 \mid x\} = \text{pr}(y = 1 \mid z_1, z_2, x)$$

which implies that the unobservable potential outcome $y(z_1, z_2)$ can be assessed through the observable outcome $y$. Then we have

$$\text{pr}\{y(z_1, z_2) = 1\} = \sum_x \text{pr}\{y(z_1, z_2) = 1 \mid x\} \text{pr}(x) = \sum_x \text{pr}(y = 1 \mid z_1, z_2, x) \text{pr}(x).$$

Inserting this into the above formulas for TE1, TE2 and INT, we obtain

$$\text{TE1} = \sum_x \text{pr}(y = 1 \mid z_1 = 1, z_2 = 0, x) \text{pr}(x) - \sum_x \text{pr}(y = 1 \mid z_1 = 0, z_2 = 0, x) \text{pr}(x), \quad (1a)$$

$$\text{TE2} = \sum_x \text{pr}(y = 1 \mid z_1 = 0, z_2 = 1, x) \text{pr}(x) - \sum_x \text{pr}(y = 1 \mid z_1 = 0, z_2 = 0, x) \text{pr}(x), \quad (1b)$$

$$\text{INT} = \left\{ \sum_x \text{pr}(y = 1 \mid z_1 = 1, z_2 = 1, x) \text{pr}(x) - \sum_x \text{pr}(y = 1 \mid z_1 = 0, z_2 = 1, x) \text{pr}(x) \right\} - \text{TE1}$$

$$= \left\{ \sum_x \text{pr}(y=1|z_1=1, z_2=1, \boldsymbol{x})\,\text{pr}(\boldsymbol{x}) - \sum_x \text{pr}(y=1|z_1=1, z_2=0, \boldsymbol{x})\,\text{pr}(\boldsymbol{x}) \right\} - \text{TE2}. \quad (1c)$$

In particular, in randomized trial, we have $\text{pr}(\boldsymbol{x}) = \text{pr}(\boldsymbol{x}|z_1, z_2)$, which implies that

$$\sum_x \text{pr}(y=1|z_1, z_2, \boldsymbol{x})\,\text{pr}(\boldsymbol{x}) = \sum_x \text{pr}(y=1|z_1, z_2, \boldsymbol{x})\,\text{pr}(\boldsymbol{x}|z_1, z_2) = \text{pr}(y=1|z_1, z_2).$$

Inserting this into (1a), (1b) and (1c), we see that TE1, TE2 and INT given by (1a), (1b) and (1c) are the same as those for randomized trials given in the introduction of this article.

### 2.3. Regression model

The risk $\text{pr}(y=1 \mid z_1, z_2, \boldsymbol{x})$ of $y = 1$ in stratum $(z_1, z_2, \boldsymbol{x})$ is modeled by logistic model. By the likelihood ratio-based significance testing of the model parameters, we obtain

$$\text{Log}\left\{\frac{\text{pr}(y=1 \mid z_1, z_2, \boldsymbol{x})}{1 - \text{pr}(y=1 \mid z_1, z_2, \boldsymbol{x})}\right\} =$$

$$\alpha + \beta_1 z_1 + \beta_2 z_2 + \beta_3 (z_1 * z_2) + \theta_1 x_1 + \theta_2 x_2 + \theta_3 x_3 + \theta_4 (z_1 * x_1). \quad (2)$$

In this model, we include, in addition to one term for the exposure product $z_1 * z_2$, another term for the hospital type-age product $z_1 * x_1$, because of a somewhat small p-value, 0.20 for the significance test of $\theta_4 \neq 0$. Let $\pi = (\alpha, \beta_1, \beta_2, \beta_3, \theta_1, \theta_2, \theta_3, \theta_4)$ be the set of all model parameters. The ML estimate $\hat{\pi} = (\hat{\alpha}, \hat{\beta}_1, \hat{\beta}_2, \hat{\beta}_3, \hat{\theta}_1, \hat{\theta}_2, \hat{\theta}_3, \hat{\theta}_4)$ and its approximate covariance matrix $\hat{\Sigma}$ (i.e. the inverse of the observed information) are given in Table 2.

### 2.4 Estimation of TE1, TE2 and INT by logistic model

From model (2), we obtain

$$\text{pr}(y=1 \mid z_1, z_2, \boldsymbol{x}) =$$

$$\frac{\exp\{\alpha + \beta_1 z_1 + \beta_2 z_2 + \beta_3(z_1 * z_2) + \theta_1 x_1 + \theta_2 x_2 + \theta_3 x_3 + \theta_4(z_1 * x_1)\}}{1 + \exp\{\alpha + \beta_1 z_1 + \beta_2 z_2 + \beta_3(z_1 * z_2) + \theta_1 x_1 + \theta_2 x_2 + \theta_3 x_3 + \theta_4(z_1 * x_1)\}} \quad (3)$$

Inserting (3) into (1a)-(1c) and replacing the probability pr($x$) by the proportion pror($x$) of the covariates $x$ in the sample, we obtain TE1, TE2 and INT as functions of $\pi$

$$\text{TE1}(\pi) = \sum_x \frac{\exp\{\alpha + \beta_1 + \theta_1 x_1 + \theta_2 x_2 + \theta_3 x_3 + \theta_4 x_1\}}{1 + \exp\{\alpha + \beta_1 + \theta_1 x_1 + \theta_2 x_2 + \theta_3 x_3 + \theta_4 x_1\}} \text{prop}(x)$$

$$- \sum_x \frac{\exp\{\alpha + \theta_1 x_1 + \theta_2 x_2 + \theta_3 x_3\}}{1 + \exp\{\alpha + \theta_1 x_1 + \theta_2 x_2 + \theta_3 x_3\}} \text{prop}(x), \quad (4a)$$

$$\text{TE2}(\pi) = \sum_x \frac{\exp\{\alpha + \beta_2 + \theta_1 x_1 + \theta_2 x_2 + \theta_3 x_3\}}{1 + \exp\{\alpha + \beta_2 + \theta_1 x_1 + \theta_2 x_2 + \theta_3 x_3\}} \text{prop}(x)$$

$$- \sum_x \frac{\exp\{\alpha + \theta_1 x_1 + \theta_2 x_2 + \theta_3 x_3\}}{1 + \exp\{\alpha + \theta_1 x_1 + \theta_2 x_2 + \theta_3 x_3\}} \text{prop}(x), \quad (4b)$$

$$\text{INT}(\pi) = \left\{ \sum_x \frac{\exp\{\alpha + \beta_1 + \beta_2 + \beta_3 + \theta_1 x_1 + \theta_2 x_2 + \theta_3 x_3 + \theta_4 x_1\}}{1 + \exp\{\alpha + \beta_1 + \beta_2 + \beta_3 + \theta_1 x_1 + \theta_2 x_2 + \theta_3 x_3 + \theta_4 x_1\}} \text{prop}(x) \right.$$

$$\left. - \sum_x \frac{\exp\{\alpha + \beta_2 + \theta_1 x_1 + \theta_2 x_2 + \theta_3 x_3\}}{1 + \exp\{\alpha + \beta_2 + \theta_1 x_1 + \theta_2 x_2 + \theta_3 x_3\}} \text{prop}(x) \right\} - \text{TE1}(\pi)$$

$$= \left\{ \sum_x \frac{\exp\{\alpha + \beta_1 + \beta_2 + \beta_3 + \theta_1 x_1 + \theta_2 x_2 + \theta_3 x_3 + \theta_4 x_1\}}{1 + \exp\{\alpha + \beta_1 + \beta_2 + \beta_3 + \theta_1 x_1 + \theta_2 x_2 + \theta_3 x_3 + \theta_4 x_1\}} \text{prop}(x) \right.$$

$$\left. - \sum_x \frac{\exp\{\alpha + \beta_1 + \theta_1 x_1 + \theta_2 x_2 + \theta_3 x_3 + \theta_4 x_1\}}{1 + \exp\{\alpha + \beta_1 + \theta_1 x_1 + \theta_2 x_2 + \theta_3 x_3 + \theta_4 x_1\}} \text{prop}(x) \right\} - \text{TE2}(\pi). \quad (4c)$$

Replacing $\pi$ in TE1($\pi$), TE2($\pi$) and INT($\pi$) by $\hat{\pi}$, we obtain the ML estimates $\widehat{\text{TE1}} = \text{TE1}(\hat{\pi})$, $\widehat{\text{TE2}} = \text{TE2}(\hat{\pi})$ and $\widehat{\text{INT}} = \text{INT}(\hat{\pi})$. Although $\hat{\pi}$ is biased for finite sample, $\widehat{\text{TE1}}$, $\widehat{\text{TE2}}$ and $\widehat{\text{INT}}$ are unbiased [17]. Clearly, $\hat{\pi}$ is consistent and so are $\widehat{\text{TE1}}$, $\widehat{\text{TE2}}$ and $\widehat{\text{INT}}$. With $\hat{\pi}$ given in Table 2, these ML estimates are equal to $\widehat{\text{TE1}} = 0.12$, $\widehat{\text{TE2}} = -0.48$ and $\widehat{\text{INT}} = 0.01$.

We are going to obtain confidence region of (TE1 or TE2, INT) by generating approximate distribution of ($\widehat{TE1}$ or $\widehat{TE2}$, $\widehat{INT}$). Let $p$ be a random variable which follows the normal distribution $N(\hat{\pi}, \hat{\Sigma})$, namely, $p \sim N(\hat{\pi}, \hat{\Sigma})$, where $\hat{\pi}$ and $\hat{\Sigma}$ in $N(\hat{\pi}, \hat{\Sigma})$ are given in Table 2. This normal distribution is good approximation to the distribution of the ML estimate $\hat{\pi}$, because parameters of a logistic model have good asymptotic normality [18]. Replacing $\pi$ in {TE1($\pi$), TE2($\pi$), INT($\pi$)} by $p$ and then using the normal distribution of $p$, we generate distribution of {TE1($p$), TE2($p$), INT($p$)}, which approximates the distribution of ($\widehat{TE1}$, $\widehat{TE2}$, $\widehat{INT}$). With this approximate distribution, we obtain approximate distribution of ($\widehat{TE1}$ or $\widehat{TE2}$, $\widehat{INT}$) and then the confidence region of (TE1 or TE2, INT).

Because {TE1($\pi$), TE2($\pi$), INT($\pi$)} is a smooth monotone bounded function of $\pi$ according to (4a)-(4c), the performance of $p$ in its approximating $\hat{\pi}$ determines the performance of {TE1($p$), TE2($p$), INT($p$)} in its approximating ($\widehat{TE1}$, $\widehat{TE2}$, $\widehat{INT}$). Exact and approximate distributions of the ML estimate of exposure effect have been simulated in the setting of a large exposure effect, a large exposure-covariate interaction and a strong confounding, and it was found that the exact and approximate distributions have rather good agreement even in their tails [17].

This method of obtaining confidence region of (TE1 or TE2, INT) is analogous to the common method of obtaining confidence interval of an odds ratio. For instance, suppose that $\beta_1$ of model (2) is a parameter of interest and we wish to obtain 95 % confidence interval of the odds ratio $\exp(\beta_1)$. The $\beta_1$ has good asymptotic normality: the distribution of $\hat{\beta}_1$ is approximately normal with a variance estimate $\widehat{\text{var}}(\hat{\beta}_1)$. Then one generates an approximate distribution of $\exp(\hat{\beta}_1)$ by using the approximate normal distribution of $\hat{\beta}_1$ and thus obtains the 95 % confidence interval of the odds ratio by $\exp\{\hat{\beta}_1 \pm 1.96 \widehat{\text{var}}^{1/2}(\hat{\beta}_1)\}$, where the number 1.96 is the 97.5$^{\text{th}}$ percentile of the standard normal distribution. This confidence interval reflects the asymmetry of the distribution of $\exp(\hat{\beta}_1)$. This method has also been used to obtain confidence intervals of other measures of exposure effect such as risk ratio [19].

In the next section, we shall describe the procedure of obtaining the approximate distribution of $(\widehat{\text{TE1}}, \widehat{\text{TE2}}, \widehat{\text{INT}})$ and various point and interval estimates of TE1, TE2 and INT.

### 3. Point and interval estimates of TE1, TE2 and INT

#### 3.1 Approximate distribution of $(\widehat{\text{TE1}}, \widehat{\text{TE2}}, \widehat{\text{INT}})$ and the derived distributions

First we draw $p$ from $p \sim \text{N}(\hat{\pi}, \hat{\Sigma})$. Second, we replace $\pi$ by $p$ in formulas (4a) - (4c) to get $\{\text{TE1}(p), \text{TE2}(p), \text{INT}(p)\}$, which approximates $(\widehat{\text{TE1}}, \widehat{\text{TE2}}, \widehat{\text{INT}})$. We iterate the procedure, 1000 times in this article, to get 1000 sets of approximate values of $(\widehat{\text{TE1}}, \widehat{\text{TE2}}, \widehat{\text{INT}})$. All these 1000 sets form an approximate distribution of $(\widehat{\text{TE1}}, \widehat{\text{TE2}}, \widehat{\text{INT}})$.

All pairs of $(\widehat{\text{TE1}}, \widehat{\text{INT}})$ in those 1000 sets form an approximate distribution of $(\widehat{\text{TE1}}, \widehat{\text{INT}})$. All $\widehat{\text{TE1}}$ in those 1000 pairs form an approximate distribution of $\widehat{\text{TE1}}$. All $\widehat{\text{INT}}$ in those 1000 pairs form an approximate distribution of $\widehat{\text{INT}}$. The three distributions are presented in Figure 1. Similarly, we obtain approximate distributions for $(\widehat{\text{TE2}}, \widehat{\text{INT}})$ and $\widehat{\text{TE2}}$, which are presented in Figure 2.

Restricting $\widehat{\text{TE1}}$ to a given range, all corresponding $\widehat{\text{INT}}$ form an approximate conditional distribution of $\widehat{\text{INT}}$ when $\widehat{\text{TE1}}$ falls into the given range. The ranges of $\widehat{\text{TE1}}$ we consider are the terciles of $\widehat{TE1}$, i.e. three consecutive ranges of $\widehat{TE1}$, such that the probability of $\widehat{TE1}$ occurring in each interval is 1 / 3. The three conditional distributions are presented in Figure 3. Similarly, we obtain three approximate conditional distributions of $\widehat{INT}$ when $\widehat{TE2}$ falls into the terciles, as presented in Figure 4.

### 3.2 Point estimate and confidence region of (TE1 or TE2, INT)

Point estimates of TE1, TE2 and INT have been obtained in Section 2.4 and are equal to $\widehat{TE1} = 0.12$, $\widehat{TE2} = -0.48$ and $\widehat{INT} = 0.01$.

In most practical studies with interval estimates, it is sufficient to have confidence regions of (TE1, INT) and (TE2, INT) rather than a confidence volume of (TE1, TE2, INT). We are going to use the distribution of ($\widehat{TE1}$, $\widehat{INT}$) obtained in Section 3.1 to obtain $(1 - \alpha)$ confidence region of (TE1, INT). We can arbitrarily choose a confidence curve that partitions the whole region of (TE1, INT) into a confidence region in which ($\widehat{TE1}$, $\widehat{INT}$) occurs with the probability $(1 - \alpha)$ and its complementary region in which ($\widehat{TE1}$, $\widehat{INT}$) occurs with the probability $\alpha$. Here we take the confidence region by imposing that the confidence curve is an ellipse and further requiring that the confidence region enclosed by the ellipse has smallest area at the $(1 - \alpha)$ confidence level.

First, we calculate $\{\text{mean}(\widehat{TE1}), \text{mean}(\widehat{INT})\}$ and the variances $\text{var}(\widehat{TE1})$ and $\text{var}(\widehat{INT})$ and the covariance $\text{cov}(\widehat{TE1}, \widehat{INT})$. Second, we use the normal distribution

$$N\left[\begin{Bmatrix}\text{mean}(\widehat{TE1})\\ \text{mean}(\widehat{INT})\end{Bmatrix}, \begin{Bmatrix}\text{var}(\widehat{TE1}) & \text{cov}(\widehat{TE1}, \widehat{INT})\\ \text{cov}(\widehat{TE1}, \widehat{INT}) & \text{var}(\widehat{INT})\end{Bmatrix}\right]$$

to obtain the $(1-\alpha)$ confidence curve for (TE1, INT) by the ellipse formula

$$\{\text{TE1} - \text{mean}(\widehat{\text{TE1}}) \quad \text{INT} - \text{mean}(\widehat{\text{INT}})\} \begin{pmatrix} \text{var}(\widehat{\text{TE1}}) & \text{cov}(\widehat{\text{TE1}}, \widehat{\text{INT}}) \\ \text{cov}(\widehat{\text{TE1}}, \widehat{\text{INT}}) & \text{var}(\widehat{\text{INT}}) \end{pmatrix}$$

$$\begin{Bmatrix} \text{TE1} - \text{mean}(\widehat{\text{TE1}}) \\ \text{INT} - \text{mean}(\widehat{\text{INT}}) \end{Bmatrix} = \chi_2^2 (1-\alpha)$$

where $\chi_2^2(1-\alpha)$ is the $100(1-\alpha)^{\text{th}}$ percentile of the central chi-square distribution with two degrees of freedom. The 95 % confidence region of (TE1, INT) is shown in Figure 1.

In the same way, we obtain the 95 % confidence region of (TE2, INT), which is presented in Figure 2.

### 3.3. Point estimates and confidence intervals of INT

We are going to use the distribution of $\widehat{\text{INT}}$ to obtain $(1-\alpha)$ confidence interval of INT. The confidence interval is not unique because the upper and lower confidence limits can be arbitrarily chosen such that the corresponding upper and lower confidence levels, denoted by $\alpha_u$ and $\alpha_l$ respectively, satisfy $\alpha_u + \alpha_l = \alpha$. Here we take the confidence interval by imposing the condition $\alpha_u = \alpha_l = 0.025$. Then the 95 % confidence interval of INT is $(-0.31, 0.34)$, which is also presented in Table 3a, together with the point estimate $\widehat{\text{INT}} = 0.01$ obtained in Section 2.4.

We are going to use the conditional distribution of $\widehat{\text{INT}}$ when $\widehat{\text{TE1}}$ falls into a terciles of $\widehat{\text{TE1}}$ to obtain point estimate and $(1-\alpha)$ confidence interval of INT over the tercile. The point estimate of INT is the mean of the conditional distribution. Over the lower tercile of $\widehat{\text{TE1}}$, the point estimate (95 % confidence interval) of INT is 0.18 (−0.05, 0.40). Over the middle tercile of $\widehat{TE1}$, the point estimate (95 % confidence interval) of INT is 0.03 (0.20, 0.21). Over the upper tercile of $\widehat{TE1}$, the point estimate (95 % confidence interval) of INT is −0.15 (−0.40, 0.04). These results are also presented in Table 3b. Similarly, over the lower, middle and upper terciles of $\widehat{TE2}$, the point estimate (95 % confidence interval) of INT are 0.18 (−0.04, 0.40), 0.01 (−0.20, 0.23) and −0.14 (−0.40, 0.14) respectively. These results are also presented in Table 3c.

### 4. Interpretation for various interval estimates of TE1, TE2 and INT

From Figures 1b, 1c and 2b, we cannot observe the correlation of $\widehat{INT}$ with $\widehat{TE1}$ or $\widehat{TE2}$. The confidence interval of INT in Table 3a indicates possible values for INT, i.e. (−0.10, 0.72), at the 95 % confidence level regardless of TE1 and TE2.

From Figure 1a, we see that $\widehat{INT}$ is highly correlated with $\widehat{TE1}$. The confidence region of (TE1, INT) indicates possible values for (TE1, INT) at the 95 % confidence level. From Figure 2a, we see similar situation for (TE2, INT).

From Figures 3a-3c, we see that INT has different values at different TE1. The confidence interval of INT over a tercile of $\widehat{TE1}$ in Table 3b indicates possible values of INT at the 95 % confidence level when $\widehat{TE1}$ falls in the tercile. Such confidence intervals allow us to perform stratified analysis of INT over strata of $\widehat{TE1}$. If the effect of large versus small hospital types for early stage cancer is $\widehat{TE1} \leq 0.105$, then we have an increase of the effect for advanced stage cancer, i.e. $\widehat{\text{INT}}$ (95 % confidence interval) = 0.18 (−0.05, 0.40). On the other hand, it is opposite situation if $\widehat{\text{TE1}} > 0.246$.

Similarly, from Figures 4a-4c and Table 3c, we see that INT has difference value at different TE2. If the effect of advanced versus early cancer stage for small hospital type is $\widehat{\text{TE2}} \leq -0.508$, then we have an increase of the effect for large hospital type, i.e. $\widehat{\text{INT}}$ (95 % confidence interval) = 0.18 (−0.04, 0.40). On the other hand, it is opposite situation if $\widehat{\text{TE2}} > -0.374$.

Although the TE2-INT relation is similar to the TE1-INT relation as revealed in the analysis above, they cannot be obtained from each other.

## 5. Discussion and conclusions

To learn the relationship between exposure effects and interaction between the exposures, one needs to report both the exposure effects and the interaction. Because the ML estimates of the exposure effects and the interaction are highly correlated, one needs to report them jointly. In this article, we have obtained point estimate and confidence region of (TE1, INT), those of (TE2, INT), and point estimate and confidence interval of INT when the ML estimate of TE1 or TE2 falls into specified range.

Because of its statistical advantages, logistic model is ubiquitous in observational studies for exposure effects on dichotomous outcomes of populations. One major advantage of a logistic model is that the model parameters have good asymptotic normality: normal distribution is good approximate distribution for the ML estimate of the model parameters, see, e.g. [18]. One major disadvantage with logistic model is the use of odds ratio as measure of the exposure effects; see the rich literature for non-collapsibility of odds ratio [20, 21, 12, 22-24]. In this article, we have kept the advantage of logistic model while avoiding the disadvantage by using risk difference as measure of the exposure effects and difference between risk differences as measure of the interaction. We have used approximate normal distribution of the ML estimate of the model parameters to obtain approximate non-normal distribution of the ML estimate of TE1, TE2 and INT and then their interval estimates including confidence regions. This maximum-likelihood-based approach provides a simple but reliable method of interval estimation of TE1, TE2 and INT, which can be easily implemented by using any software that generates normal distribution.

Two methods are available in the literature to calculate confidence interval of TE1 (or TE2 or INT in rare cases) based on logistic model [6-10], i.e. the normal approximation method and the bootstrap method, but they have not been used to calculate confidence region of (TE1, INT). In the normal approximation method, one derives approximate variance of the ML estimate of TE1 by using the delta method and then uses the variance to obtain normal approximation confidence interval of TE1. To obtain confidence region of (TE1, INT), however, one needs to derive approximate covariance matrix of the ML estimate of (TE1, INT), which is tedious. In the bootstrap method, one generates bootstrap samples by parametric or non-parametric bootstrap method and then uses the bootstrap samples to obtain bootstrap confidence interval of TE1. However, it is highly difficult to correct finite-sample bias arising from the bootstrap sampling particularly with dichotomous outcome [25--27]. It is even more difficult to correct the bias for the bootstrap confidence region of (TE1, INT) [25--27]. In comparison, our method of obtaining interval estimates including confidence regions of TE1, TE2 and INT is based on approximate distribution of their maximum estimates and does not involve resampling of the data. Therefore our method does not belong to the category of resampling methods such as the bootstrap method.

**Acknowledgement**

This research received no specific grant from any funding agency in the public, commercial, or not-for-profit sectors.

**References**

1. Blot WJ and Day NE. Synergism and interaction: are they equivalent? *American Journal of Epidemiology* 1979; **110**: 99--100.

2. Rothman KJ, Greenland S and Walker AM. Concepts of interaction. *American Journal of Epidemiology*1980; **112**: 467--470.

3. Rothman KJ, Greenland S and Lash TL. *Modern Epidemiology (3rd edition).* Lippincott Williams & Wilkins, Philadelphia; 2008

4. Greenland S. Interactions in Epidemiology: Relevance, Identification and Estimation. *Epidemiology* 2009; **20**: 14–17.

5. VanderWeele TJ. On the Distinction between Interaction and Effect Modification. *American Journal of Epidemiology* 2009; **20**: 863--871.

6. McNutt LA, Wu C, Xue X and Hafner JP. Estimating the relative risk in cohort studies and clinical trials of common outcomes. *American Journal of Epidemiology* 2009; **157**: 940--943.

7. Newcombe RG. A deficiency of the odds ratio as a measure of effect size. *Statistics in Medicine* 2006; **25**: 4235--4240.

8. Greenland S. Model-based Estimation of Relative Risk and Other Epidemiologic measures in Studies of Common Outcomes and Case-Control Studies. *American Journal of Epidemiology* 2004; **160**: 301--305.

9. Austin PC. Absolute risk reductions, relative risks, relative risk reductions, and numbers needed to treat can be obtained from a logistic regression model. *Journal of Clinical Epidemiology* 2010; **63**: 2--6.

10. Nie L, Chu H, Li F, and Cole SR. Relative excess risk due to interaction: resampling-based confidence intervals. *Epidemiology* 2010; **21**: 552--556.

11. Hansson LE, Ekstrom AM, Bergstrom R and Nyren O. Surgery for stomach cancer in a defined Swedish population: current practices and operative results. Swedish Gastric Cancer Study Group. *The European journal of surgery* 2000; **166**, 787--975.

12. Greenland S, Robins JM and Pearl J. Confounding and collapsibility in causal inference. *Statistical Science* 1999; **14**: 29–46.

13. Greenland S and Robins JM. Identifiability, exchangeability, epidemiological confounding. *International Journal of Epidemiology* 1986; **15**: 413–419.

14. Rosenbaum PR and Rubin DB. The central role of the propensity score in observational studies for causal effects. *Biometrika* 1983; **70**: 41--55.

15. Rosenbaum PR. *Observational studies*. Springer, New York; 1995.

16. Rubin DB, Wang X, Yin L and Zell E. Estimating the Effect of Treating Hospital Type on Cancer Survival in Sweden Using Principal Stratification. In *The HANDBOOK OF APPLIED BAYESIAN ANALYSIS*, eds. T. O'Hagan and M. West, Oxford University Press, Oxford; 2009.

17. Wang X, Jin Y and Yin L. Point and Interval Estimations of Marginal Risk Difference by Logistic Model. *Communications in Statistics: theory and methods* 2015; to appear.

18. Lindsey JK. *Parametric Statistical Inference*. Clarendon Press, Oxford; 1996.

19. Wang X, Jin Y and Yin L. Measuring and estimating treatment effect on dichotomous outcome of a population. *Statistical methods in medical research* 2015; doi:10.1177/0962280213502146.

20. Gail MH, Wieand S and Piantadosi S. Biased estimates of treatment effect in randomized experiments with nonlinear regressions and omitted covariates. *Biometrika* 1984; **71**: 431--444.

21. Guo J and Geng Z. Collapsibility of logistic regression coefficients. *Journal of Royal Statistical Society B* 1995; **57**: 263--267.

22. Lee Y and Nelder JA. Conditional and marginal models: another view. *Statistical Science* 2004; **19**: 219--228.

23. Austin PC. The performance of different propensity score method for estimating marginal odds ratio. *Statistics in Medicine* 2007; **26**: 3078–3094.

24. Austin PC, Grootendorst P, Normand SLT and Anderson GM. Conditioning on the propensity score can result in biased estimation of common measures of treatment effect: A Monte Carlo study. *Statistics in Medicine* 2007; **26**: 754–768.

25. Carpenter J and Bithell J. Bootstrap confidence intervals: when, which, what? A practical guide for medical statisticians. *Statistics in Medicine* 2000; **19**: 1141--1164.

26. Davison AC and Hinkley DV. *Bootstrap Methods and their Application*. Cambridge University Press, Cambridge; 1997.

27. Greenland S. Interval estimation by simulation as an alternative to and extension of confidence intervals. *International Journal of Epidemiology* 2004; **33**: 1389–1397.

**Table 1:** Descriptive statistics of the study population: one-year survivals / patient totals on levels of age, gender and geographic area for (Hospital type, Cancer stage)

|  | (Hospital type, Cancer stage) | | | |
| --- | --- | --- | --- | --- |
|  | (Large, Advanced) | (Small, Advanced) | (Large, Early) | (Small, Early) |
| **Overall** | 23/75 | 4/36 | 23/31 | 5/8 |
| **age** | | | | |
|   <= median (67) | 12/41 | 2/13 | 11/13 | 2/2 |
|   >median | 11/34 | 2/23 | 12/18 | 3/6 |
| **gender** | | | | |
|   female | 4/22 | 1/8 | 3/5 | 0/1 |
|   male | 19/53 | 3/28 | 20/26 | 5/7 |
| **Geographic area** | | | | |
|   rural | 11/32 | 4/26 | 13/15 | 3/6 |
|   urban | 12/43 | 0/10 | 10/16 | 2/2 |

**Table 2** ML estimates and its approximate covariance matrix for parameters of the model (2)

| Parameters | | $\alpha$ | $\beta_1$ | $\beta_2$ | $\beta_3$ | $\theta_1$ | $\theta_2$ | $\theta_3$ | $\theta_4$ |
|---|---|---|---|---|---|---|---|---|---|
| | Estimates | 3.92 | −3.82 | −2.63 | 0.87 | −0.06 | 0.95 | −0.66 | 0.07 |
| Covariance matrix | | | | | | | | | |
| $\alpha$ | | 12.25 | −11.68 | −1.10 | 1.08 | −0.16 | −0.37 | −0.10 | 0.16 |
| $\beta_1$ | | −11.68 | 13.25 | 1.10 | −1.36 | 0.16 | 0.04 | 0.01 | −0.18 |
| $\beta_2$ | | −1.10 | 1.10 | 0.90 | −0.90 | 0.01 | −0.00 | 0.01 | −0.01 |
| $\beta_3$ | | 1.08 | −1.36 | −0.90 | 1.14 | −0.01 | 0.02 | −0.00 | 0.01 |
| $\theta_1$ | | −0.16 | 0.16 | 0.01 | -0.01 | 0.00 | 0.00 | 0.00 | −0.00 |
| $\theta_2$ | | −0.37 | 0.04 | −0.00 | 0.02 | 0.00 | 0.27 | −0.03 | −0.00 |
| $\theta_3$ | | −0.10 | 0.01 | 0.01 | −0.00 | 0.00 | −0.03 | 0.17 | −0.00 |
| $\theta_4$ | | 0.16 | −0.18 | −0.01 | 0.01 | −0.00 | −0.00 | −0.00 | 0.00 |

**Table 3** Point estimates (50 and 95 % confidence intervals) of INT, obtained respectively from (3a) marginal distribution of $\widehat{INT}$, (3b) conditional distribution of $\widehat{INT}$ when $\widehat{TE1}$ falls into the tercile, and (3c) conditional distribution of $\widehat{INT}$ when $\widehat{TE2}$ falls into the tercile,

| | |
|---|---|
| **(3a) Point estimate (50 and 95 % confidence intervals) of INT** | |
| 0.01 (−0.31, −0.10, 0.14, 0.34) | |
| **(3b) Point estimate (50 and 95 % confidence intervals) of INT over tercile of $\widehat{TE1}$** | |
| Tercile of $\widehat{TE1}$ | |
| (−∞, 0.105] | 0.18 (−0.05, 0.10, 0.26, 0.40) |
| (0.105, 0.246] | 0.03 (−0.20, −0.03, 0.09, 0.21) |
| (0.246, +∞) | −0.15 (−0.40, −0.22, -0.08, 0.04) |
| **(3c) Point estimate (50 and 95 % confidence intervals) of INT over tercile of $\widehat{TE2}$** | |
| Trercile of $\widehat{TE2}$ | |
| (−∞, −0.508] | 0.18 (−0.04, 0.10, 0.26, 0.40) |
| (−0.508, −0.374] | 0.01 (−0.20, −0.06, 0.08, 0.23) |
| (−0.374, +∞) | −0.14 (−0.40, −0.22, −0.05, 0.14) |

**Figure 1** (1a) The scatterplot for approximate distribution of the ML estimate of (TE1, INT) and the 95% confidence region of (TE1, INT); the point estimate of (TE1, INT) is equal to $(0.12, 0.01)$. (1b) Approximate distribution for the ML estimate of INT. (1c) Approximate distribution for the ML estimate of TE1.

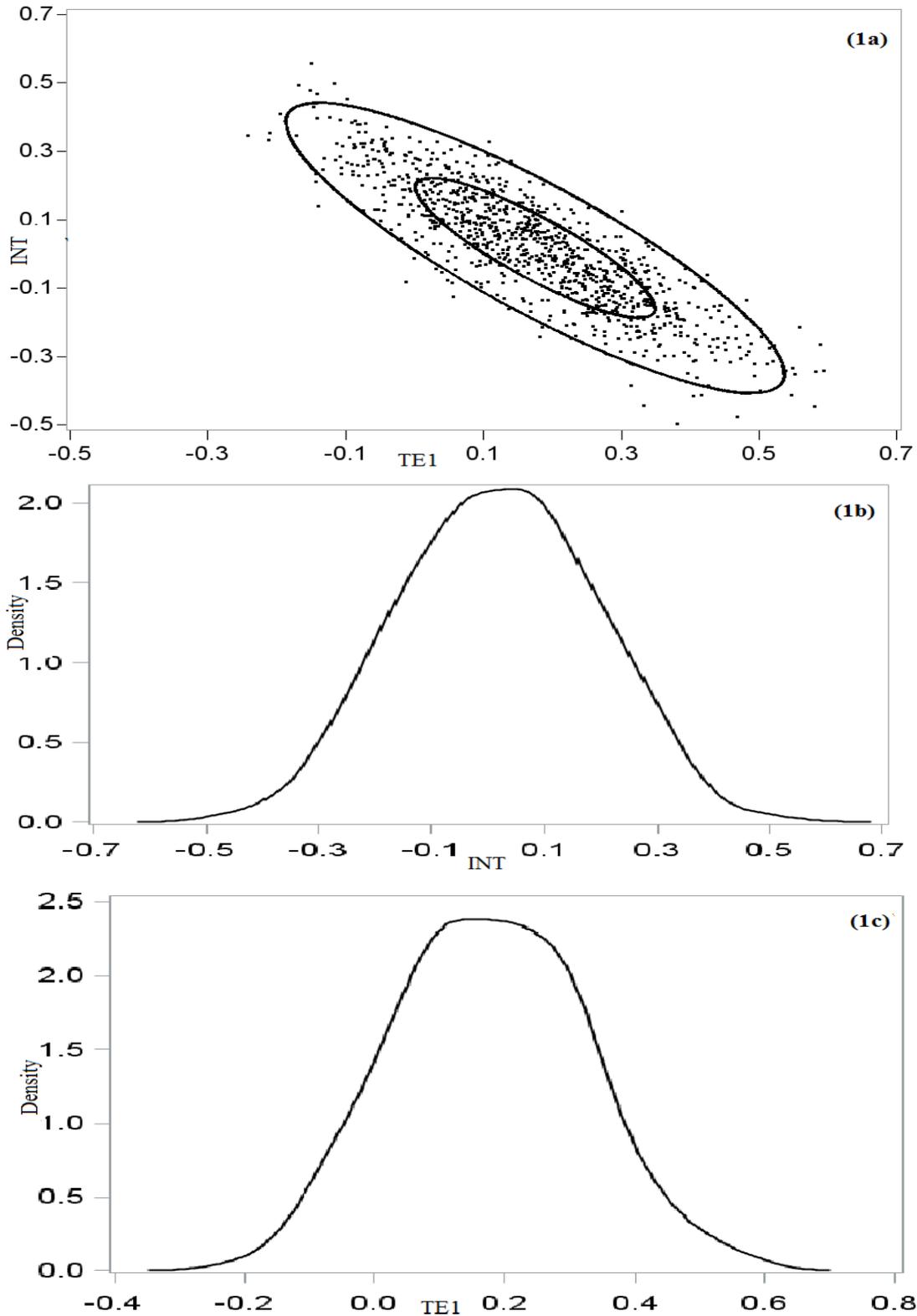

**Figure 2** (2a) The scatterplot for approximate distribution of the ML estimate of (TE2, INT) and the 95% confidence region of (TE2, INT); the point estimate of (TE2, INT) is equal to $(-0.48, 0.01)$. (2b) Approximate distribution for the ML estimate of TE2.

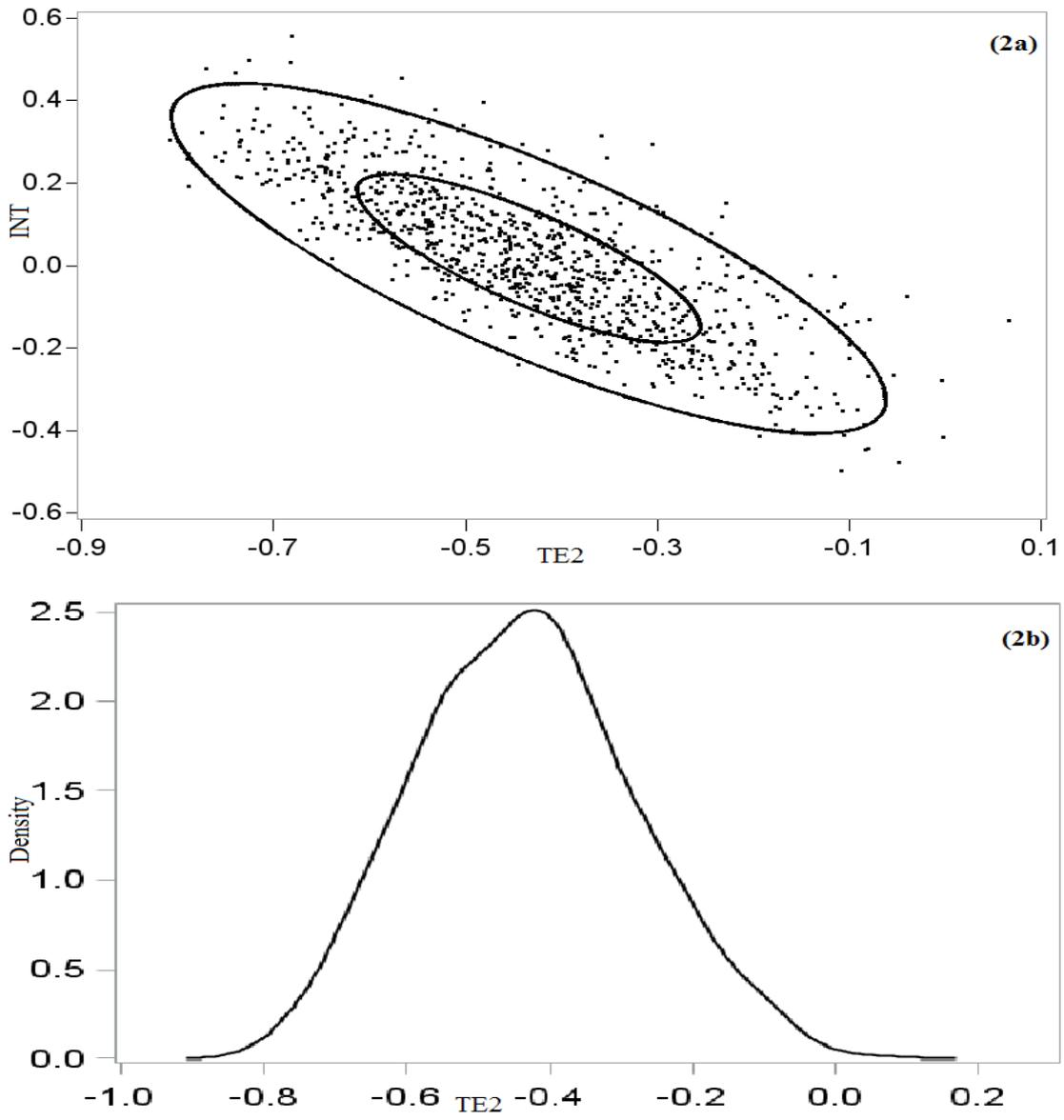

**Figure 3** Approximate conditional distribution of the ML estimate of INT when the ML estimate of TE1 falls into (3a) the lower tercile, (3b) the middle tercile, and (3c) the upper tercile.

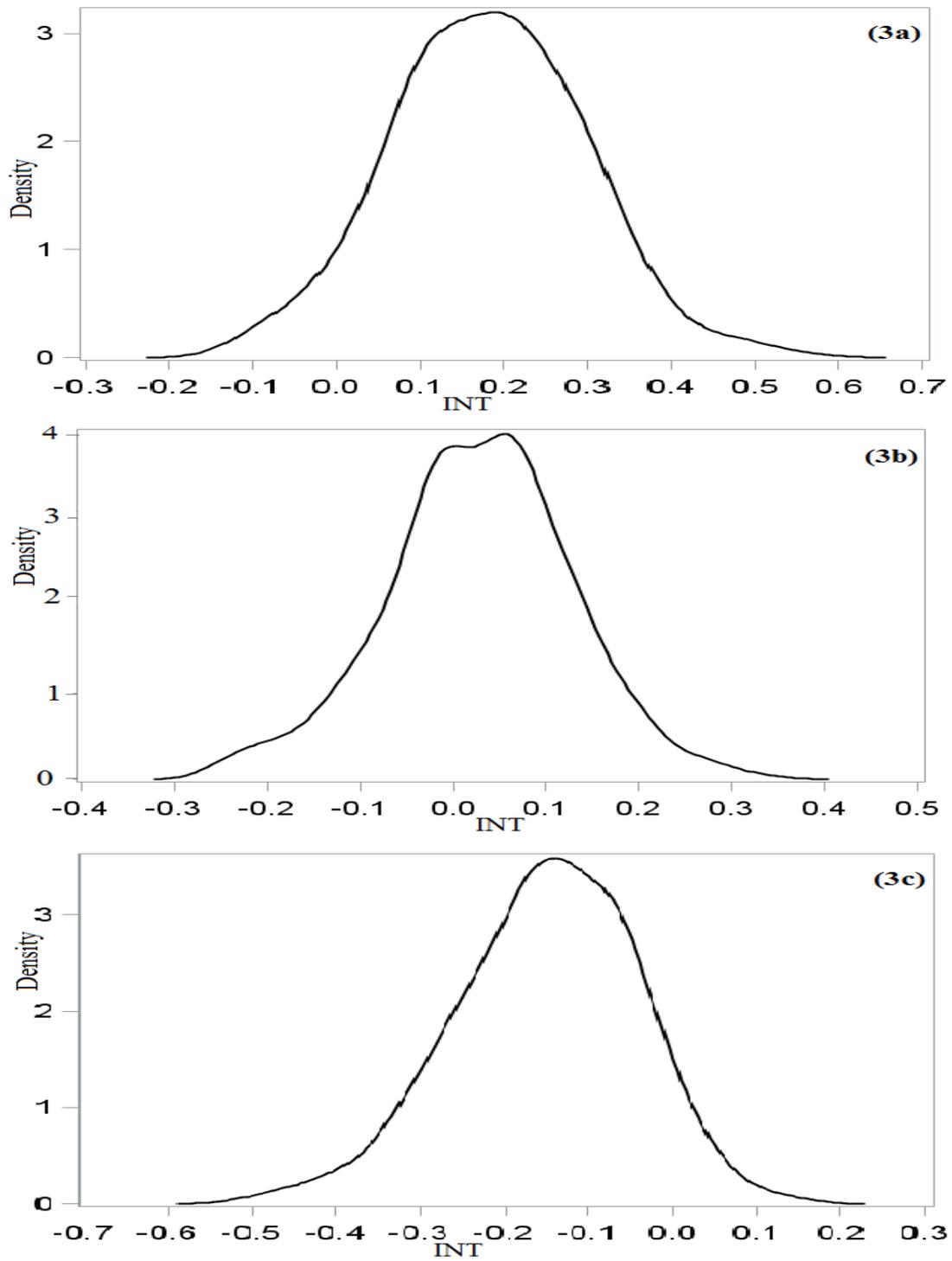

**Figure 4** Approximate conditional distribution of the ML estimate of INT when the ML estimate of TE2 falls into (4a) the lower tercile, (4b) the middle tercile, and (4c) the upper tercile.

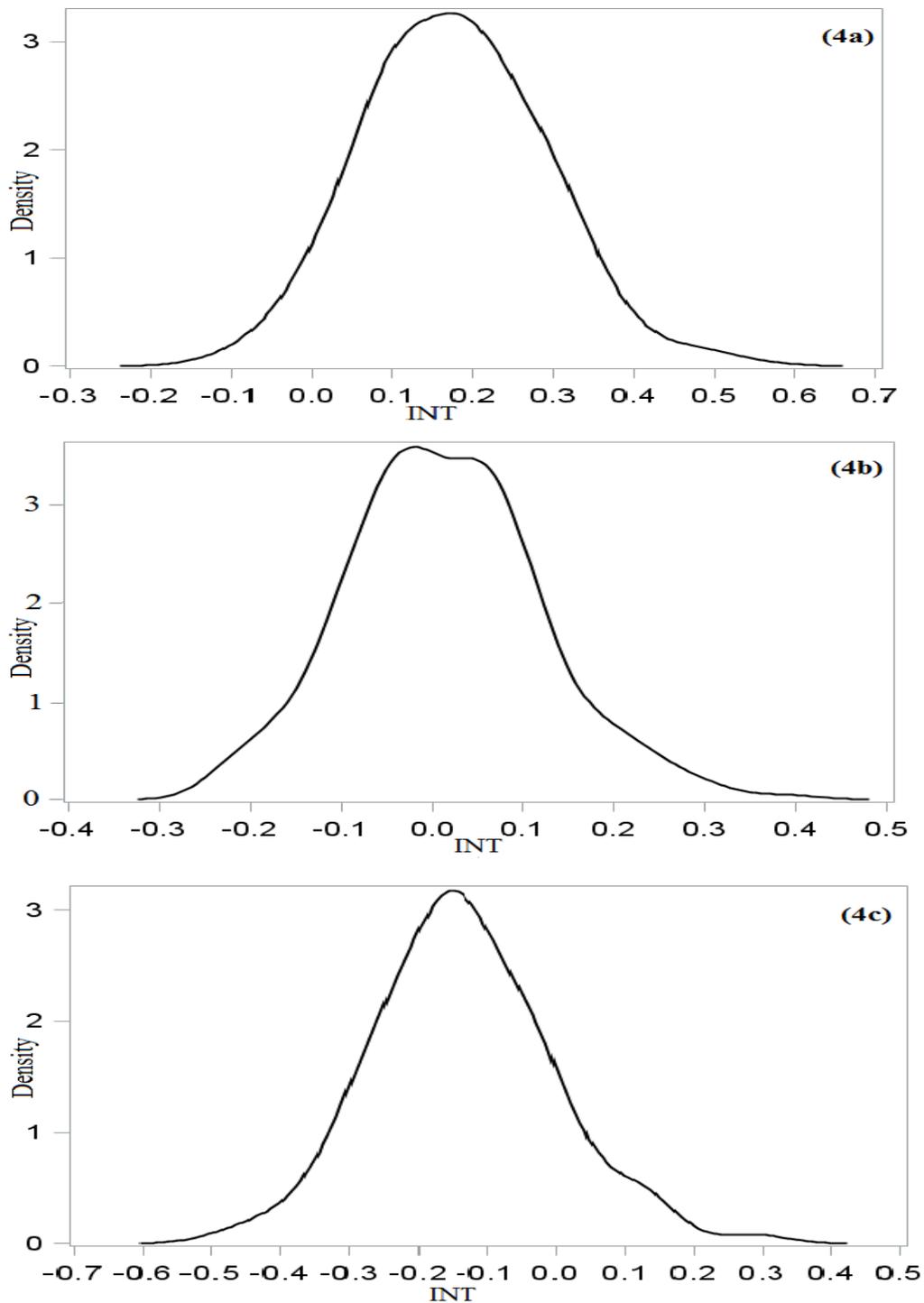